\documentclass{elsart}

\usepackage{graphicx}

\begin{document}

  \begin{frontmatter}

    \title{Bulk properties of light deformed nuclei derived from  a
           medium-modified meson-exchange interaction}

    \author[julich]{F.Gr\"ummer},
    \author[julich,ciae]{B.Q. Chen},
    \author[julich,ciae,itpb]{Z.Y. Ma} and
    \author[julich]{S.Krewald}
    \address[julich]{Institut f\"ur Kernphysik, Forschungszentrum J\"ulich GmbH,
                     D-52425 J\"ulich, Germany}
    \address[ciae]{China Institute of Atomic Energy, Beijing 102413, 
                   People's Republic of China}
    \address[itpb]{Institute of Theoretical Physics, Beijing 100080, 
                   People's Republic of China} 

    \begin{abstract}
      Deformed Hartree-Fock-Bogoliubov calculations for finite nuclei are 
      carried out. As residual interaction, a Brueckner G-matrix derived  
      from a meson-exchange potential is taken. Phenomenological medium 
      modifications of the meson masses are introduced. The binding energies, 
      radii, and deformation parameters of the  Carbon,Oxygen, Neon, and 
      Magnesium isotope chains are found to be in good agreement with the 
      experimental data.
    \end{abstract}
  
  \end{frontmatter}

In recent years, one of the most exciting discoveries in  nuclear
physics is the neutron halo and neutron skin in light unstable nuclei 
near the driplines \cite{ta85,ha87}.
The new exotic nuclei provide a challenging test for theoretical models of
nuclear structure.
As one moves away from the magic nuclei, pairing
effects become important, and the nuclear mean field may be deformed.
The effective interactions employed in many models have been adjusted
to reproduce the bulk properties of a limited set of stable nuclei,
so that the application of such interactions to nuclei in the vicinity
of the drip lines involves some extrapolations.
The presently most successful models can be grouped in two classes.

\begin{itemize}
  \item[(i)]  {The Hartree-Fock-Bogoliubov(HFB) models\cite{dob84,dob94}
               offer a combined treatment of the neutron Fermi energies and 
               the pairing gaps, so that one may expect a good prediction 
               of the neutron separation energies. HFB calculations including 
               deformations and using a G-matrix as a residual interaction 
               have been performed many years ago \cite{good,fae}, but these 
               calculations concentrated on binding energies, deformations, 
               and excitation spectra, and did not emphasize the nuclear radii. 
               In the field of exotic nuclei, HFB calculations so far have 
               relied on efffective interactions of the Skyrme type \cite{dob84}.}
  \item[(ii)] {Relativistic mean field(RMF) models\cite{ho81,ga90,sh94}
               start from meson-exchange interactions which are theoretically 
               better motivated than the Skyrme interactions. The meson coupling 
               constants are directly adjusted to the ground state properties of 
               nuclear matter and finite nuclei, however. The interaction NL-SH 
               \cite{sh93} includes the rho-meson exchange and can reproduce
               the differences between neutron and proton radii much better than
               previous interactions do. Deformation effects have been included, 
               but pairing effects are usually treated in the BCS approximation
               \cite{sh94,la95}.}
\end{itemize}

Certainly one would like to derive the effective interaction from the
bare two-nucleon interaction, but as is well-known, non-relativistic many-body 
theories starting from the bare two-nucleon interaction
so far have not been able to reproduce the saturation properties of 
nuclear matter and finite nuclei.
A relativistic extension of Brueckner theory can successfully predict
the binding energy and the density of nuclear matter,
but for finite nuclei, the problem still remains\cite{br90,mu90}.
Recently, however, some new ideas have emerged.
Brown and Rho have suggested  that the masses of mesons should drop
in a nuclear medium due to the broken scale invariance of QCD\cite{br91,ad93}.
Among other effects, a considerable influence on the saturation properties of
nuclear matter would result.
An experimental method to search for medium modifications of meson masses
is to study the production of lepton pairs in the dense and hot matter formed 
during heavy-ion collisions. Low-mass dilepton spectra have recently been
measured by the CERES Collaborators\cite{ag95}. The experiments have shown
that there is an excess of dileptons with invariant masses between
250 MeV and 1 GeV. The enhancement of dileptons in this mass region
may be explained by the modification of vector meson properties
in the nuclear medium\cite{li95,li95a,chan96}. 

At a microscopic level, there  may be many reasons for possible medium
modifications of mesons. One has to recall that the so-called sigma meson
employed in many One-Boson exchange potentials does not exist as a 
particle, but has to be interpreted as a parameterization of a correlated
two-pion exchange\cite{jack75,lohse}. 
Likewise, the omega-exchange of One-Boson exchange models might partly
parameterize a correlated rho-pion exchange\cite{jansen}.
In nuclear matter, the exchange of a correlated pion 
pair between two nucleons is modified e.g. by the Pauli blocking of the 
intermediate nucleon states and by polarizations of the propagating 
pions due to particle-hole or $\Delta$-hole excitations. These effects
would modify the properties of the effective sigma meson
in the nuclear medium.
Microscopic investigations of these effects so far have concentrated on the
modifications of the sigma and the rho meson 
\cite{aoui95}.

Given the complexity of these calculations, we feel that a more
phenomenological intermediate step might be helpful in investigating
the influence of medium modifications on the saturation problem.
For this purpose, we suggest a model. We start from a bare two-nucleon 
interaction based on a meson-exchange model\cite{ma89}
which reproduces the nucleon-nucleon scattering data and the deuteron 
properties, but allow for modifications of this interaction in the nuclear 
medium. We then calculate bulk properties of finite nuclei within the 
framework of a deformed Hartree-Fock Bogoliubov(HFB) theory.
This procedure allows us to explore a wide range of nuclei where deformation
and pairing effects may play an important role.

In order to solve the HFB equations in this model investigation limited 
to rather light nuclei, we introduce a spherical harmonic oscillator basis 
including the major oscillator shells up to $N=3$ both for protons and 
neutrons. Since in such a small basis the results will still depend on the
oscillator length b, we minimize the energy with respect to this 
parameter. A comparison with spherical Hartree-Fock calculations
performed for $^{16}O$ in a much larger basis shows that one can
expect a gain of approximately 5 MeV in the binding energy
and an increase of up to 5\% of the radii, but the qualitative
features of the calculation persist.
Time-reversal, parity, and axial symmetry are imposed on the HFB 
transformations. These symmetries restrict our model to the description of 
even-even axially deformed nuclei\cite{gruem}.
A full treatment of short range correlations in a finite basis is
extremely involved.
We therefore suggest the following approximations:
a Brueckner G-matrix in momentum space is calculated for an infinite system
with constant density. This G-matrix is then transformed to the finite
oscillator basis in order to obtain the required two-body matrix elements
of the residual interaction. Then the standard HFB equations are solved and
provide the binding energies, radii, deformation and pairing parameters
of the nuclei under consideration.
The Fermi momentum of nuclear matter was related to the average
density of the finite nucleus as follows:
\begin{equation}
  k_F = \sqrt{\frac{3}{5}}(\frac{9\pi }{8}A)^{1/3}/\sqrt{<r^2>}.
\end{equation}
Here, $A$ is the nucleon number of the isotope considered, while
$\sqrt{<r^2>}$ denotes the calculated matter rms radius.
Eq.(1) identifies the rms-radius of a sphere of nuclear matter with 
the radius $R= \sqrt{\frac{5}{3}}\sqrt{<r^2>}$
with that of the HFB wavefunction.
Selfconsistency is obtained as follows:
For a given HFB wavefunction of the finite nucleus, the rms radius is 
calculated. Now a new Fermi momentum $k_F$ can be obtained from Eq.(1),
a new G-matrix is calculated, and a new HFB wavefunction is
produced. This process has to be repeated until selfconsistency is achieved.
One should note that the present approximation does not correspond
to a local density approximation where the force in the
nuclear interior is different from that in the nuclear surface.
In the present simple model, the interaction is the same over
the whole nucleus. 
The starting energy in the G-matrix is the energy of the two initial single 
particle states.
In the present calculation, we identify the energy of the single
particle states with the average of the single particle energies obtained in
the HFB calculation. Selfconsistency with respect to the starting energy 
is demanded.
Both direct and the exchange term of the Coulomb interaction have been
taken into account.

As is well-known, in a non-relativistic framework, all the bare potentials 
BonnA, B, and C \cite{ma89} are unable to reproduce the saturation properties 
of nuclear matter, let alone finite nuclei.
In particular, the calculated radius of $^{16}O$ is much too small
for all of these potentials. An apparently simple method to consider
medium modification effects is to replace the masses of the exchanged mesons
by some reduced numbers which may be fitted to the saturation properties.
This procedure is only partially successful, however, since it turns out to 
be impossible to simultaneously reproduce both finite nuclei and nuclear
matter\cite{sc91}.  Moreover, an unrealistically strong reduction of the
meson masses is required and the results are extremely sensitive to small
variations of the meson parameters.

We therefore introduce explicitly density dependent parameters,
such as used by Hatsuda and Lee\cite{hlee}.
In a nucleus, mesons couple to nucleon particle-hole pairs. The
meson masses are therefore modified by a self-energy. For sufficiently
small densities and momentum transfers, the self energy is linear in
the Fermi momentum $k_F$. This suggests a phenomenological
modification of the meson mass proportional to $ \rho^{1/3} $. 
Some of the mesons employed in the One-Boson exchange potential
of the two-nucleon interaction partly parameterize the
correlated two-pion and the correlated rho-pion exchange and therefore
may have density dependencies stronger than $ \rho^{1/3} $. 
The Brown-Rho scaling hypothesis postulates that all meson masses
depend linearly on $\rho$. As a general ansatz, one might expand
the meson masses in a power series with respect to $ \rho^{1/3} $. 
Now one can define classes of phenomenological models by
specific truncations of the power series. In the present investigation,
we concentrate on a model which assumes a density depence of
$ \rho^{1/3} $ for the sigma meson mass and $\rho$ for the omega meson mass.
This particular choice is motivated by the observation that
the potential BonnA reproduces the binding energy of nuclear matter
reasonably well, but underbinds $^{16}O$ \cite{sc91}.
A stronger attraction at small densities therefore seems desirable.
We have checked that a medium modification of the mass of the rho-meson
has quite small influences on the saturation properties.
We therefore did not allow for variations of the rho-meson mass
in order to keep the number of free parameters as small as possible.
In this work, we choose the One-Boson exchange potential
BonnA \cite{ma89} as a bare two-nucleon interaction.
The density dependence of  the masses of the $\omega$ meson and the
two $\sigma$ mesons of the potential BonnA is parameterized
as in ref. \cite{hlee},which makes sure that for small densities,
the meson masses scale as discussed above, while for large
densities, the meson masses remain positive:

\begin{equation}
  m_\omega (\rho) = \frac{m_\omega ^0}{1+f_\omega (\rho / \rho_0)},
\end{equation}
\begin{equation}
  m_\sigma (\rho) = \frac{m_\sigma ^0}{1+f_\sigma
  (\rho / \rho_0)^\frac{1}{3}}.
\end{equation}
Here,$ m_\omega^0$ = 782.6 MeV, $m_{\sigma^1}^0$ = 550 MeV, and
$m_{\sigma^0}^0$ = 710 MeV are the mass parameters of the BonnA potential.
The values  $f_\omega = 0.036$ and $f_{\sigma^1} = 0.025$,
$f_{\sigma^0} = 0.070$ for isospin 1 and 0, 
respectively, for the modification of the two-nucleon
interaction are found by fitting the binding energy and the radius
of $^{16}O$ as well as the binding energy per particle and the
saturation density of nuclear matter.
The density of nuclear matter  $\rho_0 = 0.15 fm^{-3}$ 
has been introduced in order to make the parameters $f$ dimensionless.
In the following, we will denote  the medium-modified version of the bare 
two-nucleon interaction BonnA$^*$.
The sigma meson of the One-Boson exchange potential is an effective
meson which parameterizes the correlated two-pion exchange. In the
bare potential, different masses and coupling constants have been
assigned to the $T=0$ and the $T=1$ channels. It is therefore not
surprising that the density dependence in both channels is
different.
It has to be emphasized that the medium-dependence of the meson masses
found in the present investigation is rather model-dependent.
\\

\begin{center}
  \includegraphics[width=0.6\textwidth]{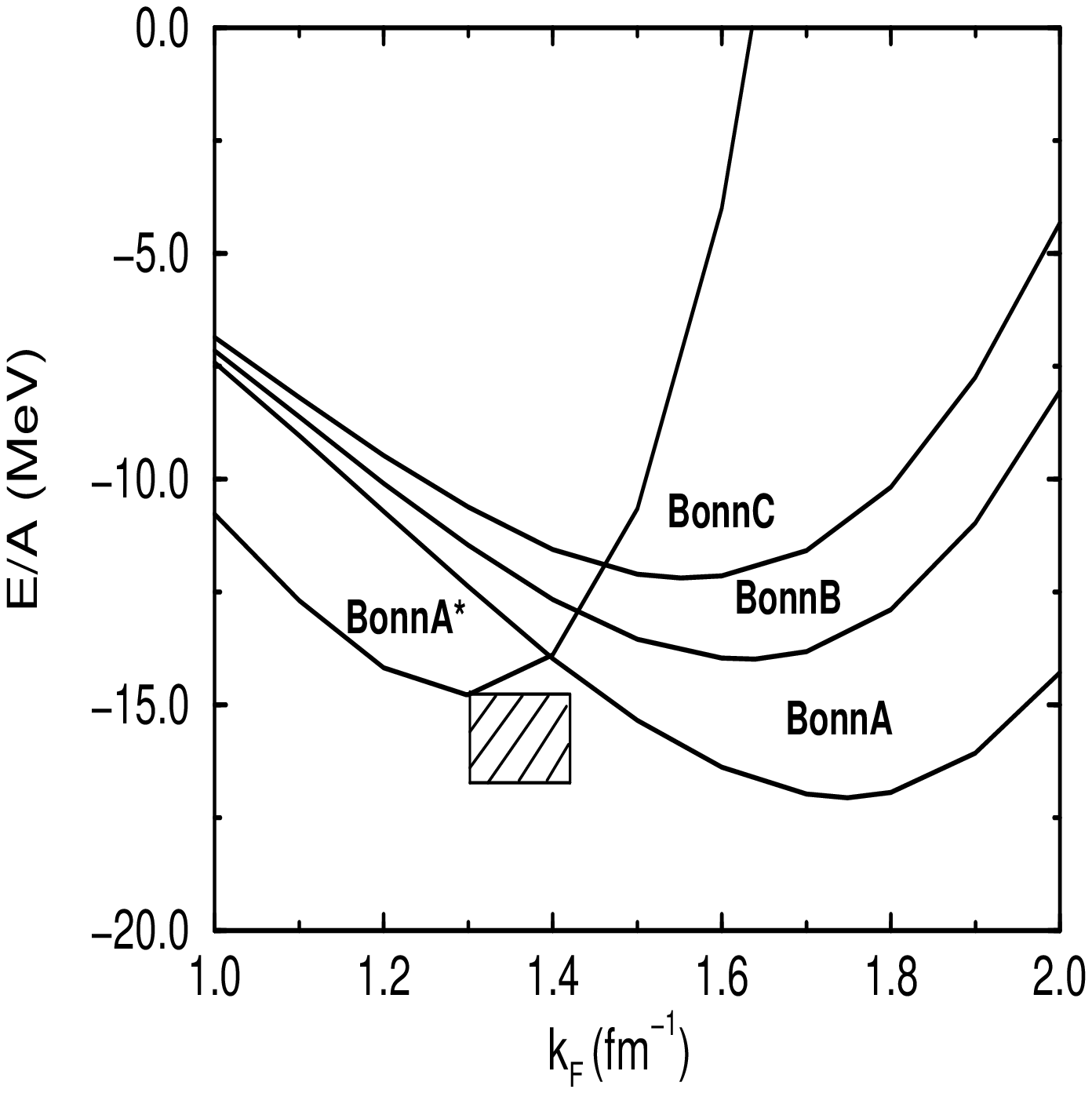}
\end{center}
Figure 1: {\it Saturation properties of nuclear matter. 
     The saturation curves are calculated in a conventional 
     BHF method with sets of One-Boson exchange potentials\cite{ma89} 
     denoted as BonnA, BonnB, BonnC and  the medium-modified
     meson-exchange potential BonnA$^*$. The shaded area shows the 
     empirical values of the saturation properties of nuclear matter.}
\\

The nuclear matter properties are calculated with  a conventional BHF 
approach. A discontinuous single particle spectrum, where the higher 
order contributions are effectively reduced, is calculated 
self-consistently. The calculated binding energy per particle is shown
as a function of the Fermi momentum in  Fig.1.  The
BonnA$^*$ potential leads to a binding energy per particle of
$ B/A = 14.6 MeV$, a Fermi momentum of $k_F=1.29 fm^{-1}$, and
a compression modulus of $K$ = 240 MeV, while $B/A$ = 16.3 MeV,
$k_F$ = 1.71 fm$^{-1}$ and $K$= 200 MeV for BonnA.

\begin{center}  
  \includegraphics[angle=270,width=0.7\textwidth]{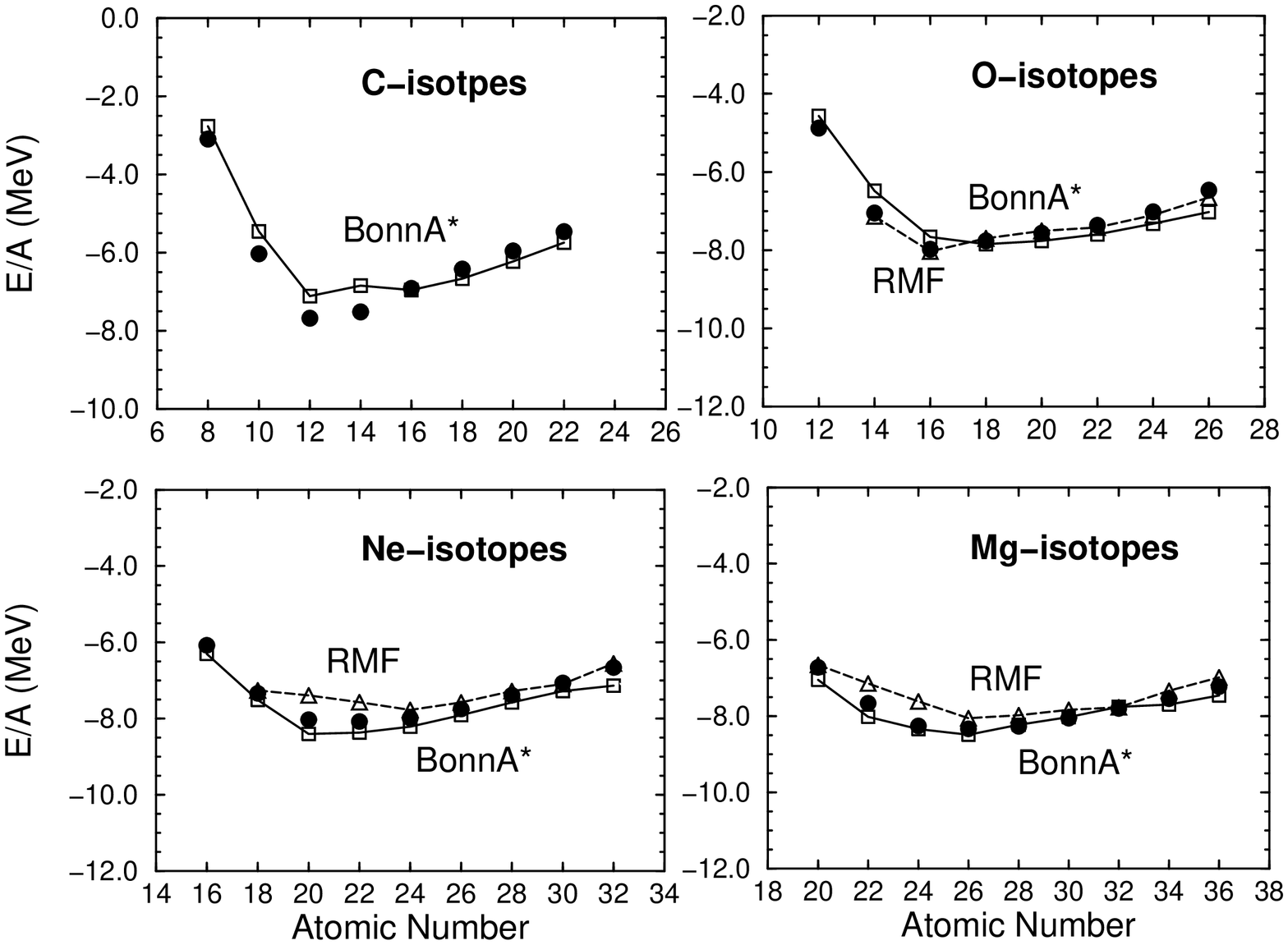}
\end{center}
Figure 2: {\it Energy per nucleon of isotope chains for light nuclei : 
     $C$, $O$, $Ne$ and $Mg$. The squares are calculated in the 
     deformed HFB approximation with a BHF G-matrix derived from 
     the medium-modified meson exchange potential BonnA$^*$. 
     The triangles are results of the spherical RMF calculations 
     with the NL-SH parameterisation. The filled solid points are 
     the experimental data\cite{au93}.}

In Fig.2, the binding energies per particle are displayed for isotopes of
the $C$, $O$, $Ne$, and $Mg$ chains. The deviation from the experimental 
binding energies per nucleon\cite{au93} is less than 0.6 MeV for all isotopes 
considered.
The slopes of the binding energy curves follow the experimental ones quite
closely. 
The importance of the nuclear deformation can be seen by comparing the
results of the present calculation with a spherical calculation, 
in particular for the isotopes
$^{20}Ne$, $^{22}Ne$ and the $Mg$-isotopes. We therefore have
performed spherical HFB calculations by forcing the deformation
to be zero and have obtained results for the isotopes mentioned above
which are very close to those obtained with the RMF-calculations.
Therefore, in the figure, we only show the RMF results.

\begin{center}   
  \includegraphics[angle=270,width=0.7\textwidth]{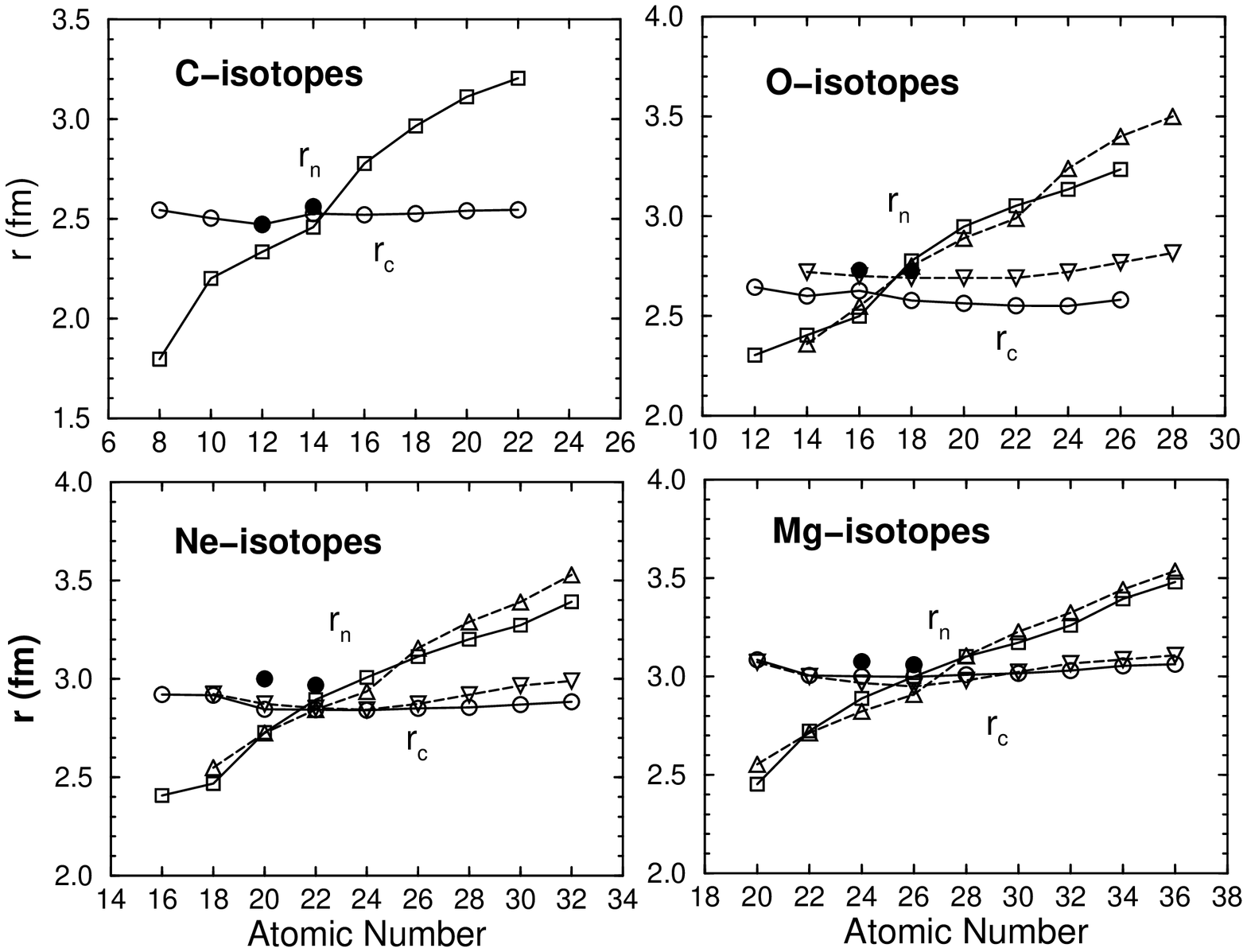}
\end{center}
Figure 3: {\it Root-mean-square radii of proton charge densities and neutron 
     point densities for the isotope chains: $C$, $O$, $Ne$ and $Mg$. 
     The squares and circles connected with solid lines are charge and 
     neutron rms radii, respectively, obtained in the deformed HFB with 
     BonnA$^*$, while the up and down triangles connected with dashed 
     lines are those from the spherical RMF calculations with the NL-SH 
     parameterization. The solid points are experimental data for charge 
     radii taken from Ref.\cite{vr87}.}

In Fig.3, the radii of the proton and neutron density distributions are shown.
The finite size of
the proton has been folded in by the prescription 
$ r_C^2=r_p^2(calculated) + r_{nucleon}^2 $
with $r_{nucleon}=0.8 fm$.
There is a fair agreement with the experimental radii  \cite{vr87}.
The present model predicts considerable neutron skins for all neutron-rich
light isotopes. The slope of the RMF calculation shows a kink between
the isotopes  $^{22}O$ and $^{24}O$, $^{24}Ne$ and $^{26}Ne$. This can be 
traced to the closure of the
$d_{5/2}$ shell. The HFB model predicts a smooth curve.

The calculated deformation parameters
for the $Ne$- and the $Mg$-isotopes are shown in Fig.4. 
They are obtained from the calculated quadrupole moments by the
relation
\begin{equation}
  \beta_2  =\sqrt{\frac{4\pi}{5}}
            \frac{A}{Z}
            \frac{Q_2^{charge}}{<r^2>_{mass}} 
\end{equation}
They are in good agreement with those from the Finite-Range Droplet 
Model (FRDM)\cite{mo95}. The dashed lines in the figure show that  
two minima for the prolate and oblate shapes are obtained in the deformed
HFB calculation. 
In the case of $^{26}Mg$, the total energy of the prolate ( $\beta_2 = 0.339$)
solution is -220.71 MeV, while the total energy of the oblate
 ($\beta_2 = -.272$)
solution is -219.63 MeV. A relatively modest change in the effective 
interaction could flip the deformation of the ground state of $^{26}Mg$
from prolate to oblate.

\begin{center}  
  \includegraphics[width=0.6\textwidth]{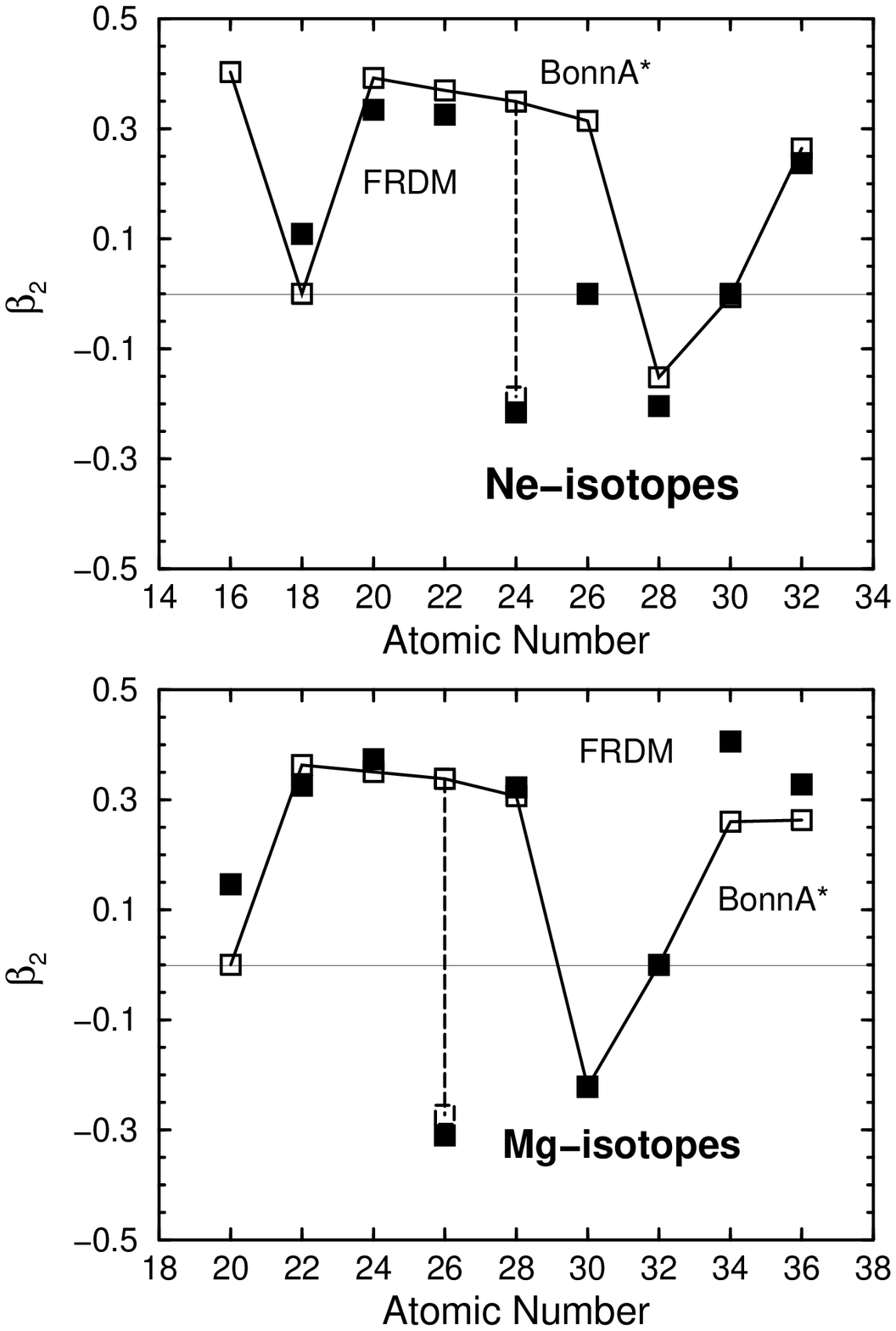}
\end{center}
Figure 4: {\it Deformations for Ne- and Mg-isotopes. The notations are same as in
     Fig.2. The filled squares are the results of the Finite-Range 
     Droplet Model (FRDM)\cite{mo95}. The dashed lines connect two 
     minima of the prolate and oblate deformations.}

To conclude, the results of Figs.1-4 show that a density-dependent modification
of the meson masses of a bare two-nucleon interaction is sufficient to 
reproduce the saturation properties of both nuclear matter and finite nuclei.
Despite the relatively small number of  free parameters, 
the agreement with the experimental data is comparable to the one
achieved with more phenomenological interactions.
We find that the deformation parameters of the nuclei may be quite
sensitive to the details of the saturation mechanism.
The  results of the present study are of such quality that we feel
encouraged  to apply this model also to isotope chains in regions of heavier
masses. The density dependence of other meson parameters should be studied
systematically. We believe that the present results justify the effort
to develop and apply microscopic theories of the medium modifications
of all low-mass mesons.

    This work was supported by the Deutsche Forschungsgemeinschaft(DFG)
and China National Natural Science Foundation. B.Q.C and Z.Y.M are 
grateful for the hospitality of Institut f\"ur Kernphysik at the
Forschungszentrum J\"ulich during their visit. The authors thank J.Speth
for valuable discussions and his constant support.

\end{document}